\documentclass[a4paper,fleqn]{cas-sc}
\usepackage{subcaption}

\usepackage{enumerate}
\usepackage{cas-common}
\usepackage{graphicx}
\usepackage{nicefrac}
\usepackage{float}
\usepackage{placeins}
\usepackage{afterpage}
\usepackage{tikz}
\newcommand*\circled[1]{\tikz[baseline=(char.base)]{
            \node[shape=circle,draw,inner sep=1pt] (char) {#1};}}

\usepackage[numbers]{natbib}

\xspaceaddexceptions{]}
\def\tsc#1{\csdef{#1}{\textsc{\lowercase{#1}}\xspace}}
\tsc{WGM}
\tsc{QE}
\tsc{EP}
\tsc{PMS}
\tsc{BEC}
\tsc{DE}

\restylefloat{figure}

\begin{document}

\makeatletter
\renewcommand\section{\@startsection{section}{1}{\z@}%
    {15pt \@plus 3\p@ \@minus 3\p@}%
    {4\p@}%
    {
     \sectionfont\raggedright\hst[13pt]}}
\renewcommand\subsection{\@startsection{subsection}{2}{\z@}%
    {10pt \@plus 3\p@ \@minus 2\p@}%
    {.1\p@}%
    {
     \ssectionfont\raggedright }}
\renewcommand\subsubsection{\@startsection{subsubsection}{3}{\z@}%
    {10pt \@plus 1\p@ \@minus .3\p@}%
    {.1\p@}%
    {
     \sssectionfont\raggedright}}
\makeatother

\let\WriteBookmarks\relax
\def\floatpagepagefraction{1}
\def\textpagefraction{.001}
\shorttitle{NectarCAM detectors performances}
\shortauthors{Tsiahina et al.}
\graphicspath{{Images/}}
\title [mode = title]{Measurement of performance of the NectarCAM photodetectors}                      



\author[1]{A. Tsiahina}[type=editor,
                        auid=000,bioid=1,
                        prefix=,
                        role=,
                        orcid=
                        ]
\cormark[1]
\fnmark[1]
\ead{atsiahina@irap.omp.eu}

\author[1] {P. Jean}
\author[1] {J.-F. Olive}
\author[1] {J. Kn\"odlseder}
\author[1] {C. Marty}
\author[1] {T. Ravel}
\author[1] {C. Jarnot}

\author[2] {B. Biasuzzi}
\author[3] {J. Bolmont}
\author[4] {F. Brun}
\author[3] {S. Caroff}
\author[5] {E. Delagnes}
\author[6] {S. Fegan}
\author[6] {G. Fontaine}
\author[7] {D. Gascon}
\author[4] {J.-F. Glicenstein}
\author[8] {D. Hoffmann}
\author[3] {S. Karkar}
\author[3] {J.-P. Lenain}
\author[7] {J. Paredes}
\author[9] {P.-O. Petrucci}
\author[10] {J. Prast}
\author[7] {M. Rib\'o}
\author[11] {S. Rivoire}
\author[7] {A. Sanuy}
\author[2] {P. Sharma}
\author[2] {T. Suomijarvi}
\author[12] {L.A. Tejedor}
\author[3] {F. Toussenel}
\author{for NectarCAM}
\author{CTA}

\address[1] {Institut de Recherche en Astrophysique et Plan\'etologie, CNRS-INSU, Universit\'e Paul Sabatier, 9 avenue Colonel Roche, BP 44346, 31028 Toulouse Cedex 4, France}
\address[2] {Laboratoire de Physique des 2 infinis, Irene Joliot-Curie,IN2P3/CNRS, Universit\'e Paris-Saclay, Universit\'e de Paris, 15 rue Georges Clemenceau, 91406 Orsay, Cedex, France}
\address[3] {Sorbonne Universit\'e, Universit\'e Paris Diderot, Sorbonne Paris Cit\'e, CNRS/IN2P3, Laboratoire de Physique Nucl\'eaire et de Hautes Energies, LPNHE, 4 Place Jussieu, F-75005 Paris, France}
\address[4] {IRFU, CEA, Universit\'e Paris-Saclay, B\^at 141, 91191 Gif-sur-Yvette, France}
\address[5] {IRFU/DEDIP, CEA, Universit\'e Paris-Saclay, B\^at 141, 91191 Gif-sur-Yvette, France}
\address[6] {Laboratoire Leprince-Ringuet, \'Ecole Polytechnique (UMR 7638, CNRS/IN2P3, Institut Polytechnique de Paris), 91128 Palaiseau, France}
\address[7] {Departament de F\'isica Qu\`antica i Astrof\'isica, Institut de Ci\`encies del Cosmos, Universitat de Barcelona, IEEC-UB, Mart\'i i Franqu\`es, 1, 08028, Barcelona, Spain}
\address[8] {Aix Marseille Univ, CNRS/IN2P3, CPPM, 163 Avenue de Luminy, 13288 Marseille cedex 09, France}
\address[9] {Univ. Grenoble Alpes, CNRS, IPAG, 414 rue de la Piscine, Domaine Universitaire, 38041 Grenoble Cedex 9, France}
\address[10] {LAPP, Univ. Grenoble Alpes, Univ. Savoie Mont Blanc, CNRS-IN2P3, 9 Chemin de Bellevue - BP 110, 74941 Annecy Cedex, France}
\address[11] {Laboratoire Univers et Particules de Montpellier, Universit\'e de Montpellier, CNRS/IN2P3, CC 72, Place Eug\`ene Bataillon, F-34095 Montpellier Cedex 5, France}
\address[12] {EMFTEL department  and IPARCOS, Universidad Complutense de Madrid, 28040 Madrid, Spain}


\begin{abstract}[S U M M A R Y]
\noindent NectarCAM is a camera for the medium-sized telescopes of the Cherenkov Telescope Array (CTA), which covers the energy range of 100 GeV to 30~TeV. The camera is equipped with 265 focal plane modules (FPMs). Each FPM comprises 7 pixels, each consisting of a photo-multiplier tube, a preamplifier, an independently controlled power supply, and a common control system. We developed a dedicated test bench to validate and qualify the industrial FPM production and to measure the performance of each FPM in a dark room before its integration in the camera. We report the measured performance of 61~FPM prototypes obtained with our experimental setup. We demonstrate that the gains of the photo multiplier tubes are stable and that pulse widths, transit time spreads, afterpulse rates and charge resolutions are within the specifications for NectarCAM.

\end{abstract}

\maketitle

\section{Introduction}

The Cherenkov Telescope Array (CTA) Observatory will provide a leap forward in sensitivity in the 0.02-300~TeV gamma-ray domain compared to the existing imaging atmospheric Cherenkov telescopes (IACT) such as H.E.S.S., VERITAS and MAGIC. The CTA Observatory will improve our understanding of particle acceleration processes in the Universe through the observation of very-high-energy gamma-ray sources (such as active galactic nuclei, supernova remnants, pulsar wind nebulae, see \cite{CTAScience:2017}). It will also allow us to probe the interstellar and intergalactic media and to explore physics beyond the standard model by searching for signatures of dark matter or effects of quantum gravity.

The CTA Observatory will be composed of Cherenkov telescopes of three sizes. These telescopes will be installed at the Instituto de Astrof\'{i}sica de Canarias (IAC) site on La Palma (Canary Islands, Spain) and at the European Southern Observatory (ESO) site at Paranal (Chile), with up to eight large-sized telescopes (LST; 4 at La Palma, and 4 at Paranal), 40 medium-sized telescopes (MST; 15 at La Palma and 25 at Paranal), and 70 Small-Sized Telescopes (SST; all at Paranal). CTA is about to complete its pre-construction phase, and will soon enter the construction phase. A prototype LST was constructed at the La Palma site and is currently under commissioning. Completion of the array is foreseen in 2025, and the science operations are expected to last over 30 years.
CTA will use the IACT technique to image atmospheric air showers produced by gamma and cosmic rays when interacting with the Earth's atmosphere. For general reference about the CTA project and the IACT techniques, see \cite{ExpAstron:2011}.

Two types of cameras are currently developed for the MSTs: NectarCAM \cite{NectarCAM:2015} and FlashCam \cite{FlashCAM2012, FlashCAM2017}. Both cameras are optimized for the detection of gamma-ray showers in the 0.1 - 30 TeV energy range with a field of view of 8$^\circ$ and a pixel size of 0.18$^\circ$. NectarCAM cameras are designed, developed, tested and commissioned by a consortium of 16 European institutes under French leadership. Among the participating institutes, the Institut de Recherche en Astrophysique et Plan\'etologie (IRAP) laboratory is responsible for the development and provision of the focal plane instrumentation composed of photomultiplier tubes (PMTs) and their associated electronics, providing the power, the signal amplification and the control. The focal plane is divided into 265 independent modules, each equipped with seven PMTs and the electronics to drive them. In mid-2018, a batch of 61 prototype modules were produced. These have been extensively tested with a dedicated test bench developed and built at IRAP. The aim of these tests was to validate the functioning of the modules produced by an industrial partner, verify that their performance meets the NectarCAM specifications, and to perform an initial calibration before their integration into the NectarCAM camera at Institut de Recherche sur les lois Fondamentales de l'Univers (IRFU), Saclay. The tests also allowed us to evaluate the test bench itself for later use by an industrial partner, where it will serve for validation of mass-produced modules in the CTA construction phase.

In section 2, we describe the design of a NectarCAM focal plane module (FPM). The design of our test bench, which allows functional tests and performance measurements on seven FPMs in parallel, is described in the next section. In section 4, we present the methods and the results of the most important performance tests obtained with the 61 focal plane module prototypes, namely: the stability of PMT gains, the charge resolution, the single photo-electron pulse shape (width and transit-time spread) and the afterpulse rate. The measured performance is compared with previous measurements and discussed in the context of the CTA requirements \cite{MSTReq:LevelA, MSTReq:LevelB, HintonMSTReq14}.

\section{The Focal Plane Module of NectarCAM}\label{Text:FPM_Nect}
\begin{figure}[H]
  \centering
  \includegraphics[width=7cm]{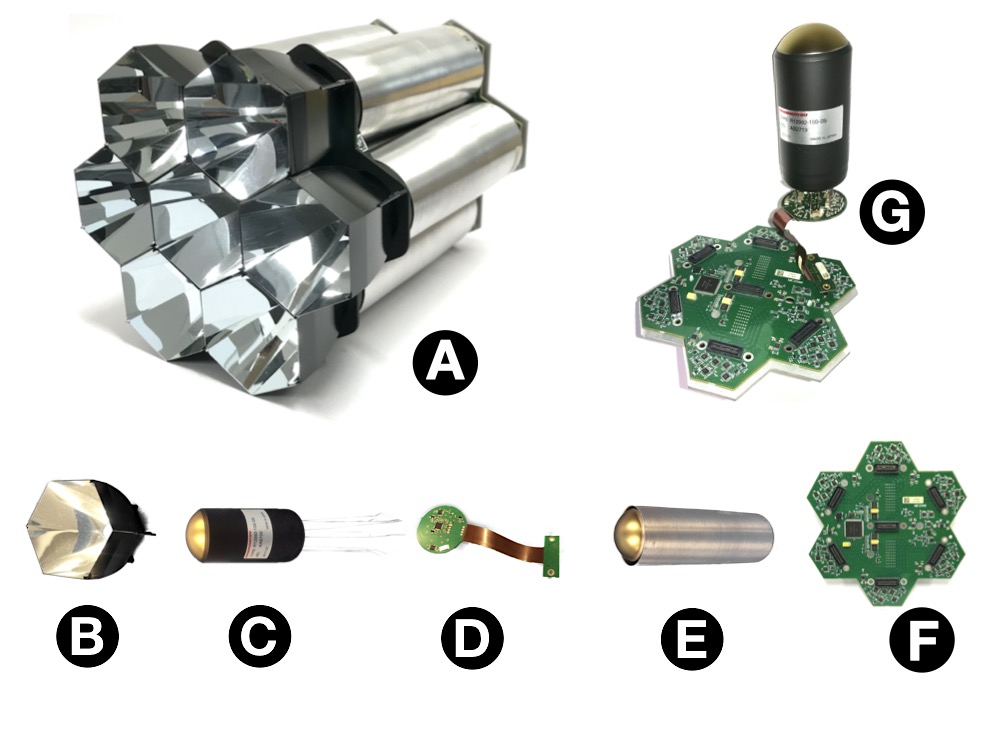}
  \caption{
  Overview of a Focal Plane Module (FPM).
  \protect\circled{A} is the FPM equipped with 7 Winston Cones (WC), \protect\circled{B} is a WC, \protect\circled{C} a PMT from Hamamatsu, \protect\circled{D} is a HVPA board, \protect\circled{E} is a Detector Unit (DU), \protect\circled{F} is the Interface Board (IB) and \protect\circled{G} is a PMT soldered to a HVPA board and connected to the IB.}
  \label{FIG:FPM_overview}
\end{figure}

NectarCAM has a total of 1855 pixels, sensitive to individual photons, organised in 265 individual FPMs (see \circled{A} in Figure \ref{FIG:FPM_overview}). Each FPM contains seven detector units (DU, \circled{E}), connected to an interface board (IB, \circled{F}, \circled{G}) equipped with a micro-controller. Each DU is composed of a PMT (\circled{C}) soldered to a High-Voltage Pre-Amplifier board (HVPA, \circled{D}) and enclosed in a grounded metal tube. The detectors are seven-dynode PMTs (R12992-100 from Hamamatsu \cite{HamamatsuPMTDataSheet}) specially designed for CTA in order to optimize the performance of the CTA imaging cameras (see \cite{Toyama15}). These PMTs have a good quantum efficiency (up to $40$\%), a high photo-electron collection efficiency, a short pulse width, a short transit time spread and low after-pulse probability. Finally, for astrophysical observations, Winston cones (WC, \circled{B}) are placed in front of each DU as a light guide in order to maximize the light collection at the photocathode. Since the aim of our tests is to characterise the FPMs, we do not use the WCs in our test bench; they are tested independently at Institut de Plan\'etologie et d'Astrophysique de Grenoble (IPAG) laboratory (see \cite{Hen13} and \cite{Hen17}), and the full NectarCAM detector chain is subsequently tested at IRFU.

The HVPA boards that are soldered to every PMT provide high voltage to the dynodes and the photocathode through a Cockcroft-Walton (CW) type electronic circuit. The CW provides a high voltage of up to 1550~V. It has been designed so that the high voltage stability (offset and ripple) is smaller than 1~V to ensure that the PMT gain fluctuations will not affect the charge resolution. In addition to providing high voltage to the PMT, the HVPA also amplifies the anode signal through a dedicated PACTA (pre-amplifier for CTA) ASIC developed by the University of Barcelona (see \cite{Sanuy2012}). In order to accommodate a dynamic range from ~1 up to ~3000 photo-electrons, PACTA splits the anode signal into two lines: a high-gain (HG) line with typical trans-impedance of $1200~\Omega$ with a maximum deviation of $\pm 60~\Omega$ and a low-gain (LG) line with typical trans-impedance of $79~\Omega$ with a maximum deviation of ${+3}/{-5}~\Omega$. The mean and standard deviation of the HG/LG ratio have been measured to be 15 and 2, respectively.

Each HVPA is connected to the IB that sets the seven desired high voltages with a closed-loop control system. The IB also routes the HG and LG signals to the front-end electronic board (FEB) on $100~\Omega$ impedance lines. Furthermore, the IB digitalizes the load currents (total current of the HVPA) and the CW current consumption. The latter is a measure of the anode current and, therefore, of the photocathode illumination. The CW current is compared via an analog circuit on the IB to a pre-defined, tunable threshold level in order to protect the PMTs in case of high levels of illumination. Such high levels may arise, for example, if a very bright star passes through the field of view, or if the camera is accidentally exposed to light. If the CW current exceeds the defined threshold, the IB will automatically shut down the PMT high voltage in less than $\sim$~100~ms. The IB communicates with the FEB via a SPI protocol. The FEB remotely controls the IB in order to get/set slow control data such as PMTs voltages, ramp speed voltage (high voltage increment per second), status, HVPA current consumption, CW current consumption and CW threshold currents.

Each FPM is connected to a FEB provided by Laboratoire de Physique Nucl\'eaire et de hautes \'Energies (LPNHE). The FEB functionalities include the triggering system of the camera \footnote{The FEB generates the trigger signals that will be transferred to the digital trigger board in order to elaborate the camera trigger.}, final amplification of the signal through an ACTA (amplifier for CTA) ASIC (see \cite{Sanuy2012b}), analog to digital conversion of amplified anode signals, data packet assembly and transmission to the computer that controls the module. The readout system of the FEB is based on the NECTAr chip, which performs the sampling, storage of the analog signal and digitization of two differential lines with a 12-bit Analog to Digital Converter (ADC) and a sampling rate from 0.5 to 2~GHz (see \cite{Naumann2012}). For each triggered event, the FEB delivers 14 waveforms which are the digitized anode currents of each HG/LG lines sampled with a nominal period of 1~ns (i.e. a sampling rate of 1~GHz). The FEBs were developped by IRFU, LPNHE, Laboratoire Univers et Particules de Montpellier (LUPM), Laboratoire de Physique Subatomique et de Cosmologie (LPSC) and Institut de Ci\`encies del Cosmos, University of Barcelona (ICCUB).

\section{The FPM test bench}\label{Text:FPM_TB}

\begin{figure}[H]
  \centering
  \includegraphics[width=15cm]{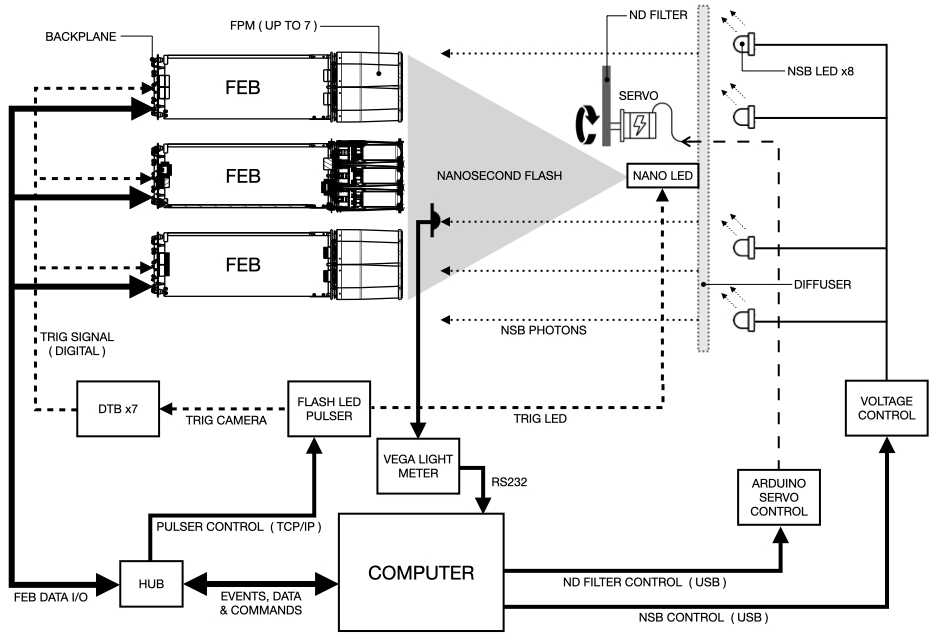}
  \caption{The FPM test bench diagram.}
  \label{FIG:FPM_TB_diag}
\end{figure}

\begin{figure}[H]
\centering

\begin{subfigure}[t]{.4\textwidth}
\centering
\includegraphics[height=5.5cm]{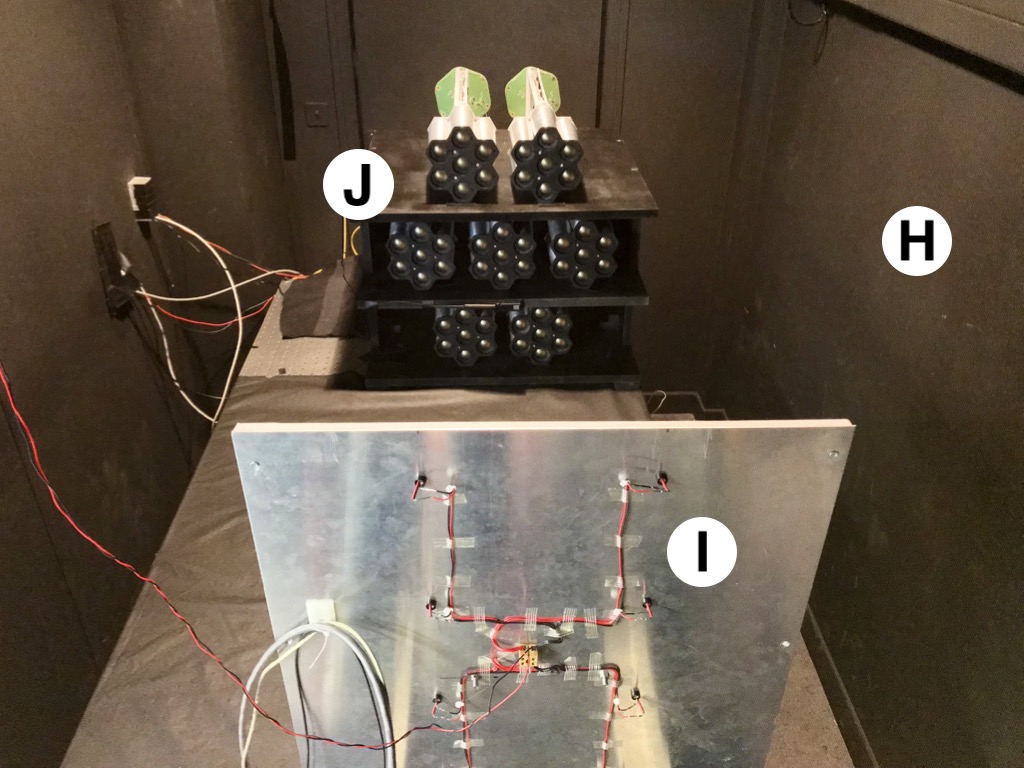}
        \caption{}\label{fig:fig_a}
\end{subfigure}
\begin{subfigure}[t]{.4\textwidth}
\centering
\includegraphics[height=5.5cm]{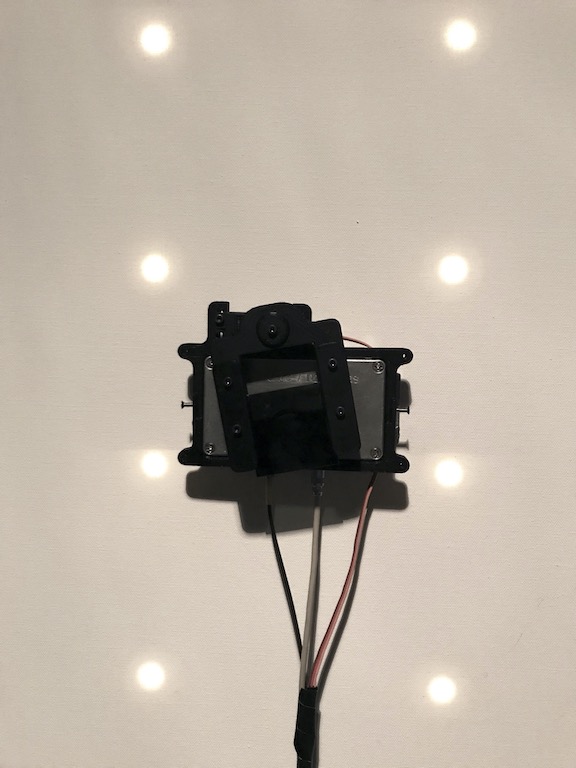}
\caption{}\label{fig:fig_b}

\end{subfigure}
\caption{Picture of the FPM test bench. In (\subref{fig:fig_a}) \protect\circled{H} is the dark room at IRAP laboratory, the frame \protect\circled{I} is the light generation system and \protect\circled{J} is the module rack. In (\subref{fig:fig_b}) is the other side of the frame \protect\circled{I} with the 8 white NSB LEDs fully powered and at the center the flash LED covered by the motorized ND filter.}
\label{FIG:FPM_TB_pic}
\end{figure}

The FPM test bench is composed of a light generation and a readout system which are connected to the FPM under test (see Figure \ref{FIG:FPM_TB_diag}). Both systems are controlled by a central computer. The light generation system is composed of a LED (called nano LED hereafter) that produces nanosecond duration UV flashes ($\lambda = 400 {~\rm nm}$) and a set of LEDs emitting continuous white light to simulate the night sky background (NSB) noise. The nanosecond flashes are produced by a LED (HERO HUVL400-510) that is driven by a LED pulser system designed by LPNHE. The light intensity is adjustable by a 8-bit DAC (with a non-linear scale); the measured light pulse intensity resolution is $\sim$~4\%. We add a controlled motorized neutral-density filter in front of the nano LED to have an extended dynamic range in pulse intensity, in particular to produce flashes at the level of a single photo-electron (ph.e.) per PMT. Without the filter, the maximum flash intensity (obtained with the LED pulser tuned to its maximum value) produces $\sim$ 5000 ph.e. in the PMT photocathode. With the filter, the minimum flash intensity (obtained with the LED pulser tuned to its minimum value) produces $\sim$ 0.03 ph.e. The attenuation factor of the filter is 12.5 \footnote{The attenuation factor was measured with an optical spectrometer and verified with PMTs using photon pulses.}. The flash LED pulser provides two analog triggering signals: one to drive the LED and the other one to trigger the readout system. The delay between the two analog signals is set remotely in order to fine tune the signal position in the readout window relative to the corresponding trigger signal. 
Eight white LEDs, behind a diffuser frame (see \circled{I} in Figure \ref{FIG:FPM_TB_pic}) and powered by a controlled DC power supply, simulate the NSB with rates ranging from 0.003 to 15~ph.e./ns produced in PMT photocathodes. The NSB level is monitored by a Vega light meter (Vega model 7Z01560 from Ophir Photonics) equipped with a NIST photodiode sensor installed in the plane of the FPM photocathodes. The Vega light meter is controlled by the computer via an RS232 port. 

We use seven NectarCAM FEBs as the readout system of our test bench, allowing us to test seven FPMs (49 PMT DUs) simultaneously. Each FEB amplifies, samples and digitizes the anode signals of FPMs (see section \ref{Text:FPM_Nect}) and transmits the resulting waveforms to the computer. The backends of the FEBs are connected to NectarCAM digital trigger board backplanes (DTB) that link the FEBs to the computer via ethernet cables and a hub (see \cite{Schwanke15}). The backplanes also transmit the external trigger signal produced by the flash LED pulser to the FEBs. The seven DTBs, produced by Deutsches Elektronen-Synchrotron (DESY), Zeuthen, convert the analog triggering signal to digital triggering signals which are delivered to the FEBs. The trigger frequency theoretically ranges up to 156~kHz although we are limited by the communication protocols to 500~Hz for measurements with 7~FPMs. 

The computer communicates with the FEB via UDP and TCP/IP protocols. It receives the digitized pulse currents of each PMT and the slow control data (see section \ref{Text:FPM_Nect}). The computer sends control commands to the FEB to manage waveform acquisitions and to tune the high voltage of each PMT. It also controls the flash LED via a TCP/IP protocol, the filter wheel via an Arduino and the DC power that supplies the eight white LEDs via USB connection. 

Our FPM test bench is installed in a dark room ($1.8\times 3.6$~${\rm m^2}$) with a controlled average air temperature of 18~$^{\circ}$C with an uncertainty of 2~${^\circ}$C (see Figure \ref{FIG:FPM_TB_pic}). The distance between the light generation system and the photocathodes plane is 103~cm. The computer that controls all the elements of the test bench with Python 2.7 scripts is installed outside the dark room. 

\section{Measurements}\label{Text:Meas}

In order to validate the proper functioning of the FPMs after their assembly and prior to their integration into NectarCAM we perform a series of measurements on each of them using the IRAP test bench. These measurements verify that their performance meets a series of NectarCAM targets concerning, for example, the PMT gain, the stability of the high-voltage supply, the signal-to-noise ratio, and the afterpulsing rate. The measurements also yield a series of ``start-of-life'' calibrations for each DU, which can then be used as a reference throughout its lifetime; these include the initial gain versus voltage curve, afterpulsing fraction, single photo-electron spectrum, and impulse response. In this section, we report the measurement methods and the results of the tests used to verify the NectarCAM detector properties with our experimental setup. These results are compared with those presented in previous studies and with the NectarCAM requirements.

\subsection{PMT gain measurements, tuning and stability with NSB rate}\label{Text:Gain}
The PMT gain is the multiplication factor of one photo-electron charge. CTA plans to operate the PMTs with a nominal gain of 40000 in order to guarantee a detector lifetime that is compliant with an observatory lifetime of 30 years while achieving a sufficient signal-to-noise ratio that still allows the detection of individual photo-electrons. CTA requires the systematic uncertainty in the determination of the gain to be less than 3\% (see \cite{Brown2015}). A first series of tests therefore ensures that a gain of 40000 can be achieved for each PMT within the allowable voltage range for its CW supply, and that the gain is sufficiently stable over the range of night-sky background photon rates expected in operation.

\begin{figure}[H]
  \centering
  \includegraphics[width=7cm]{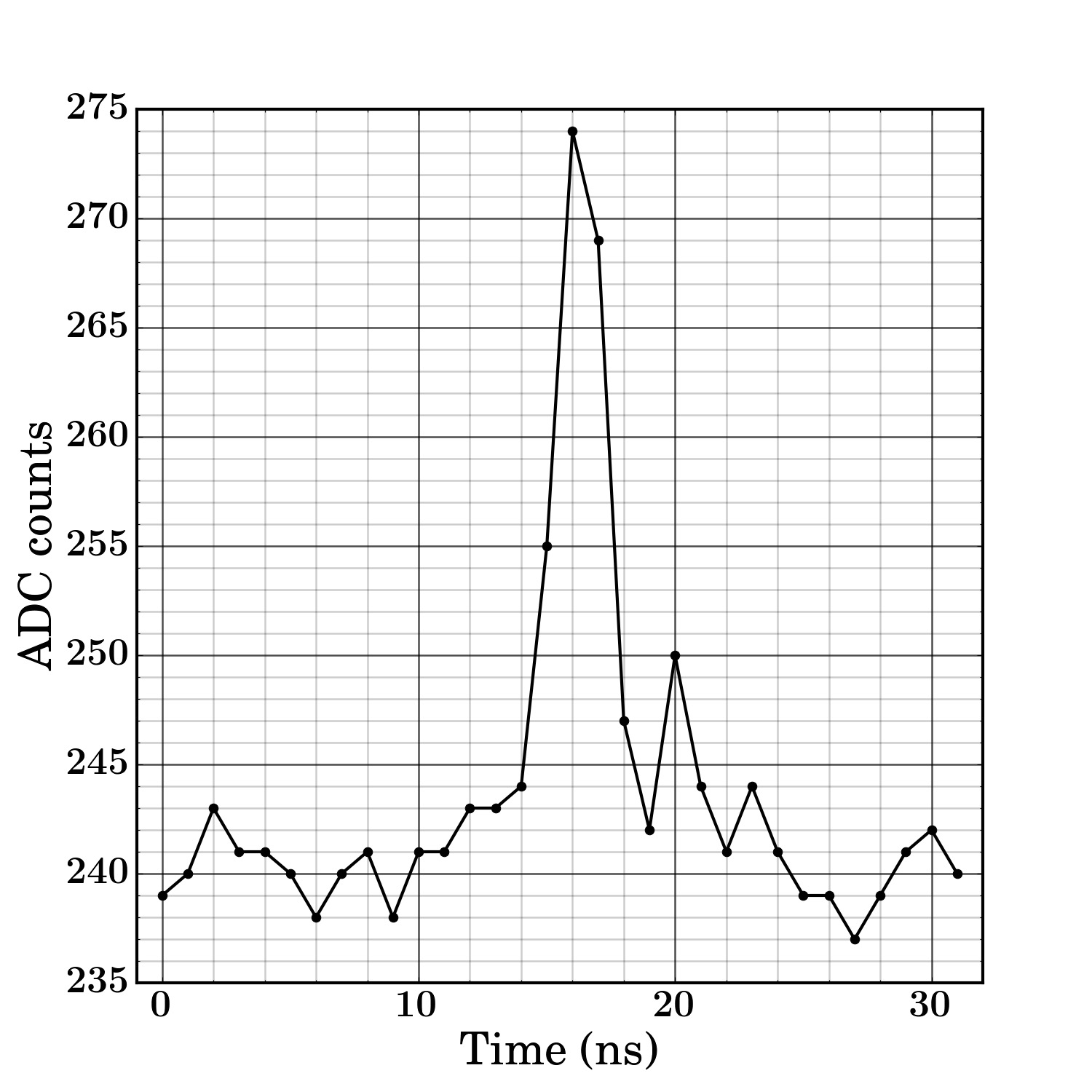}
  \caption{Example of a digitized single photo-electron anode current pulse measured with the FPM test bench.}
  \label{FIG:real_spe_plot}
\end{figure}

\begin{figure}[H]
  \centering
  \includegraphics[width=8cm]{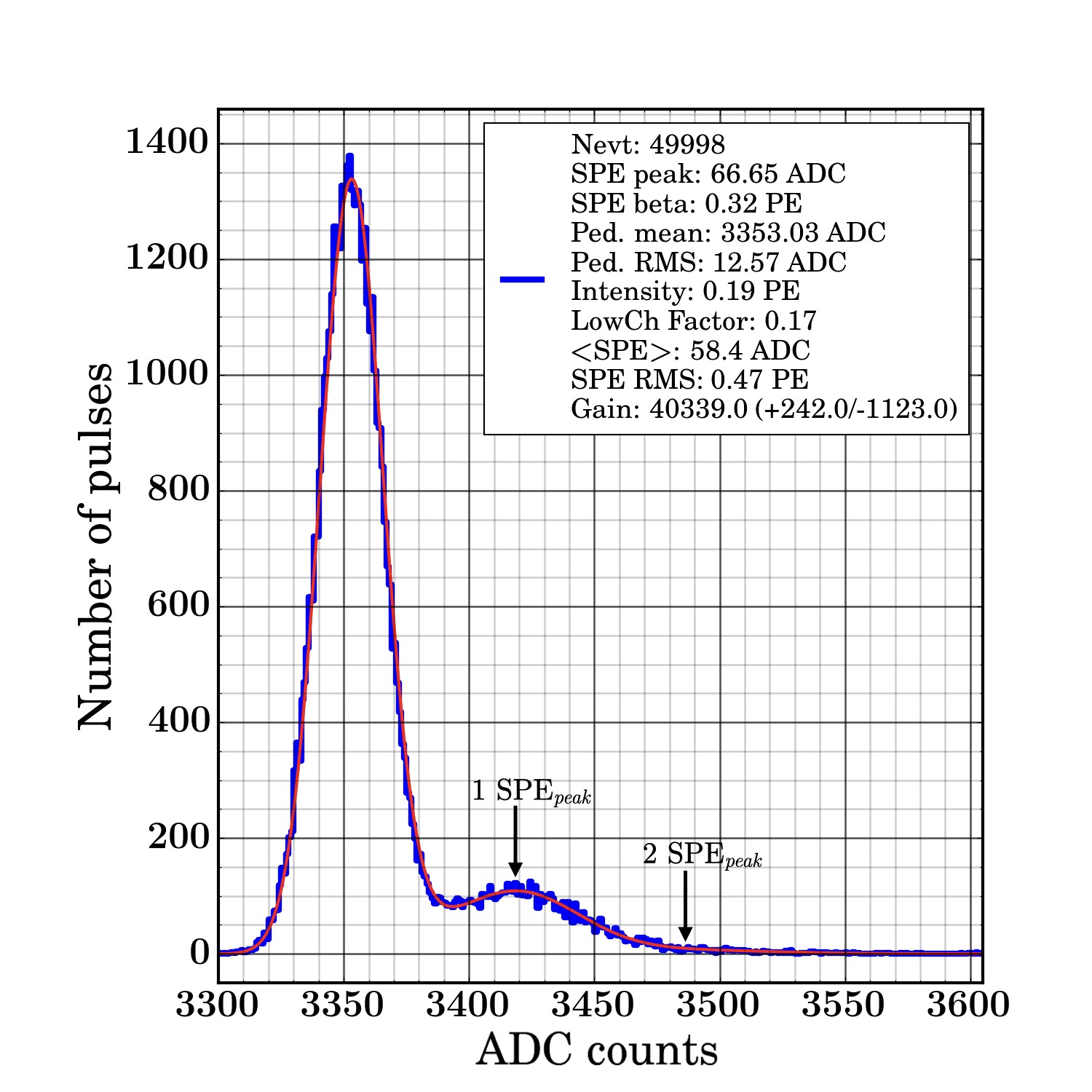}
  \caption{Example of a charge spectrum recorded with 50000 events. The blue curve shows the measured spectrum, the red curve shows the best fitting model. The best fitting parameters are included as a legend.}
  \label{FIG:myspefitter_plot}
\end{figure}

Several methods are used to measure the gain of PMTs for Cherenkov telescopes (e.g. see \cite{Bellamy94}, \cite{Hanna08}, \cite{FeganReport}, \cite{Takahashi18}). The method adopted in the present study is the single photo-electron fitting which is widely used in high-energy astrophysics instrumentation. Its main advantage is that, regardless of the flash LED pulser metrological characteristics (i.e. thermal stability, flash fluctuations), the number of photo-electrons can be determined in the fit. During operations, with NectarCAMs mounted on MST structures, an integrated single photo-electron calibration system will measure the gains with the same method (see \cite{Biasuzzi20}). 
The single photo-electron fitting method is based on fitting the position in ADC counts of the single photo electron peak in charge spectra accumulated at low luminosity of incident light.
The charges are obtained by integrating the anode currents in the time window where the pulses are expected. Since the anode currents are digitized (see section \ref{Text:FPM_Nect}), the charges are estimated with a simple sum of the samples in the given time window (see Figure \ref{FIG:real_spe_plot}). The charge unit is measured in ADC counts. It contains the baseline, also called pedestal (which fluctuates due to electronic noise) and the possible photo-electron induced pulse contributions. The latter contribution to the charge is called the net charge; it is estimated by subtracting the baseline value from the measured charge.
For illustration, a typical charge spectrum is shown in Figure \ref{FIG:myspefitter_plot}. The first peak, situated around 3350 ADC counts, is produced by events without any photo-electron; its mean value represents a programmable "pedestal" offset, while its width is determined by the noise level of the PMT and its acquisition chain. The second peak, at around 3420 ADC counts, corresponds to the case that a single photo electron was produced in the photo cathode due to the incidence of a photon. This second peak is called the Single Photo-Electron (SPE) peak. A third peak, corresponding to two photo-electrons, expected at 3490 ADC counts, is not seen in Figure \ref{FIG:myspefitter_plot} because the source luminosity is very low.

The measurement, with the FPM test bench, is performed in a low-intensity regime of the pulsed nano-LED, such that one photo-electron is produced in the photocathode for every five light flashes, on average. The number of collected events is 50000 and the charges are estimated by integrating the anode signal in a 14~ns time window (see section \ref{Text:PulseShape}). Before starting the measurement, FPMs are warmed up for about one hour to ensure stability of the electronics and PMT responses. 
The gain is derived by fitting the charge spectrum to the analytical model adopted by \cite{Biasuzzi20} and \cite{Caroff19}, using a downhill simplex method (Nelder-Mead). The analytical model is the same as the one used by \cite{Biasuzzi20} and \cite{Caroff19}, for the NectarCAM project. It consists of a Gaussian, parametrized by its centroid and its width ($\sigma_{ped}$), to account for the pedestal peak and a single photo-electron distribution convolved with a Poisson law to account for the distribution of the number of photo-electrons produced per light pulse. The single photo-electron distribution is based on a Gaussian centered to the SPE peak and a half Gaussian\footnote{The normalized distribution of a half Gaussian centered on the origin is given by ${f(y)={\frac {\sqrt {2}}{\sigma {\sqrt {\pi }}}}\exp \left(-{\frac {y^{2}}{2\sigma ^{2}}}\right)\quad y\geq 0}$ and $f(y) = 0$ when $y < 0$. $\sigma$ is the standard deviation of the normal distribution.\\} to account for the low-charge component. This model is suggested by simulations of photomultiplier response to single photo-electron  (\cite{Anthony18}). The half Gaussian could be attributed to events with a lower electron multiplication at the first dynodes, such as when the photo-electron is produced at the first dynode instead of at the photocathode or when the photo-electron misses the first dynode or when electrons are backscattered from the dynode (see \cite{Bellamy94}, \cite{Takahashi18} and \cite{KL12}). The fraction of the single photo-electron events in the half Gaussian is called the low charge fraction. In our analyses, the low-charge fraction is fixed to a value of 0.17 per single photo-electron. This value is an average derived from measurements at high gain with few PMTs (see the conclusion for further discussions of the low-charge component fraction). The width of the low charge component distribution is related to the gain and SPE resolution using the relationship described in \cite{Caroff19}. The single photo-electron distribution is convolved with a normal distribution with a width of $\sigma_{ped}$ to account for the charge dispersion due to electronic noise fluctuations. A sample spectrum and its best-fit parameters are shown in Figure \ref{FIG:myspefitter_plot}. Among the fitted parameters of the model, we obtain the mean single photo-electron charge which is the mean value of the best fitting single photo-electron distribution relative to the mean pedestal\footnote{A charge value relative to the mean pedestal is called a net charge.}. Considering the camera measurement chain (signal pre-amplification by the PACTA, line impedances and amplification by the ACTA), a single photo-electron amplified with a gain of 40000 corresponds to a mean net charge of $\bar{Q}_{SPE} \equiv$ 58 ADC counts. The other parameters provided by the fit are the pedestal characteristics (mean and standard deviation $\sigma_{ped}$), the single photo-electron rms and the intensity (mean number of photo-electrons per light pulse). The total number of free parameters is 7.

A PMT gain of 40000 (nominal gain) is obtained by tuning the high voltage such that the mean single photo-electron charge is 58 ADC. In such a condition, the high voltage is called the nominal high voltage. In principle this nominal high voltage could be found iteratively by taking SPE spectra, fitting the parameters, and adjusting the voltage using some ad-hoc gain vs voltage function. However, the large number of required events (the spectrum is made with 50000 events to get a best fit of the mean single photo-electron charge with a statistical uncertainty of $\sim$~0.5\%) and the computing time of the model fit (around 1 minute per spectrum in our i7 8 core computer) prevent us from using this method for a large number of FPMs due to the long acquisition and analysis durations. The high voltage tuning method has to be fast because, during the mass production phase, FPMs have to be calibrated quickly along the production flow to reduce the operation costs in industry. Instead, we implemented an iterative two-step method that uses the mean charge produced by intense light pulses (measurement in a high-intensity regime) as a tracer of the gain variation as a function of high voltage. The intensity of the flash LED pulser is tuned such that the mean number of photo-electrons is about 60 ph.e. per event. With such an intensity, the signal from the HG line is strong enough while staying far from saturation (at $\sim$ 200 ph.e.). The mean charge is measured with 4000 events by integrating the anode pulses in a 16~ns time window. In this condition the accuracy of the resulting mean net charge is better than 0.5\%. We estimate the gain assuming that the mean net charge is the sum of individual photo-electron charges, i.e. the gain is the measured mean net charge divided by the known number of photo-electrons. 
In a first step, the number of photo-electrons produced in each PMT photocathode is derived from three first acquisitions performed at an initial high voltage: (i) a single photo-electron spectrum measured in the low-intensity regime to estimate the initial PMT gain, (ii) a mean charge measurement in the high-intensity regime that produces about 60 ph.e. per light pulse and (iii) a measurement without light pulse to estimate the electronic pedestal with 4000 events. The mean net charge $\bar{Q}_{init}$ is obtained by subtracting from the mean charge the mean electronic pedestal value and the number of photo-electrons is the mean net charge divided by the initial gain $G_{init}$. With those values, we derive $\bar{Q}_{nom}$, the mean net charge expected when the gain is 40000 : $\bar{Q}_{nom} = \bar{Q}_{init} \times 40000 / G_{init}$.
The second step is an iterative process that tunes the high voltage to the nominal value $HV_{nom}$ until the measured mean net charge reaches the $\bar{Q}_{nom}$ value. This step consists of consecutive high-intensity and pedestal acquisitions (as the previous cases (ii) and (iii)) at several high voltages. The change in high voltage between each iteration is based on the observation that the gain varies as a power-law of the high voltage (e.g. see Figure 7 of \cite{Toyama15} and Figure 4 of \cite{Mir17}). The iterations are stopped once the difference between the measured mean net charge and the $\bar{Q}_{nom}$ target value is less than 0.5\%. This accuracy limit is determined by our precision limit of the high-voltage measurement on the IB that was designed to have a setting and measurement accuracy of better than 1~V. Finally the last step consists of measuring the photo-electron spectrum with 50000 events in the low-intensity regime in order to fit the final gain with its uncertainty as well as other parameters which characterize the performance of the PMT (mean and rms values of the electronic noise and single photo-electron rms). 
This method is faster than the interpolation method in the SPE regime. We need only a few thousand events to get a reliable mean net charge value. However its drawback is that we have to know the mean number of photo-electrons with an initial single photo-electron fit so the error of the initial gain measurement propagates through the other measurements. The nominal voltages are all between 920~V and 1150~V, well within the range that can be generated by the CW supply (see section \ref{Text:FPM_Nect}), leaving considerable room to increase the voltages during the lifetime of the PMTs to compensate for gain decrease with time, or should CTA wish to take data with a higher gain value.

\begin{figure}[H]
  \centering
  \includegraphics[width=8cm]{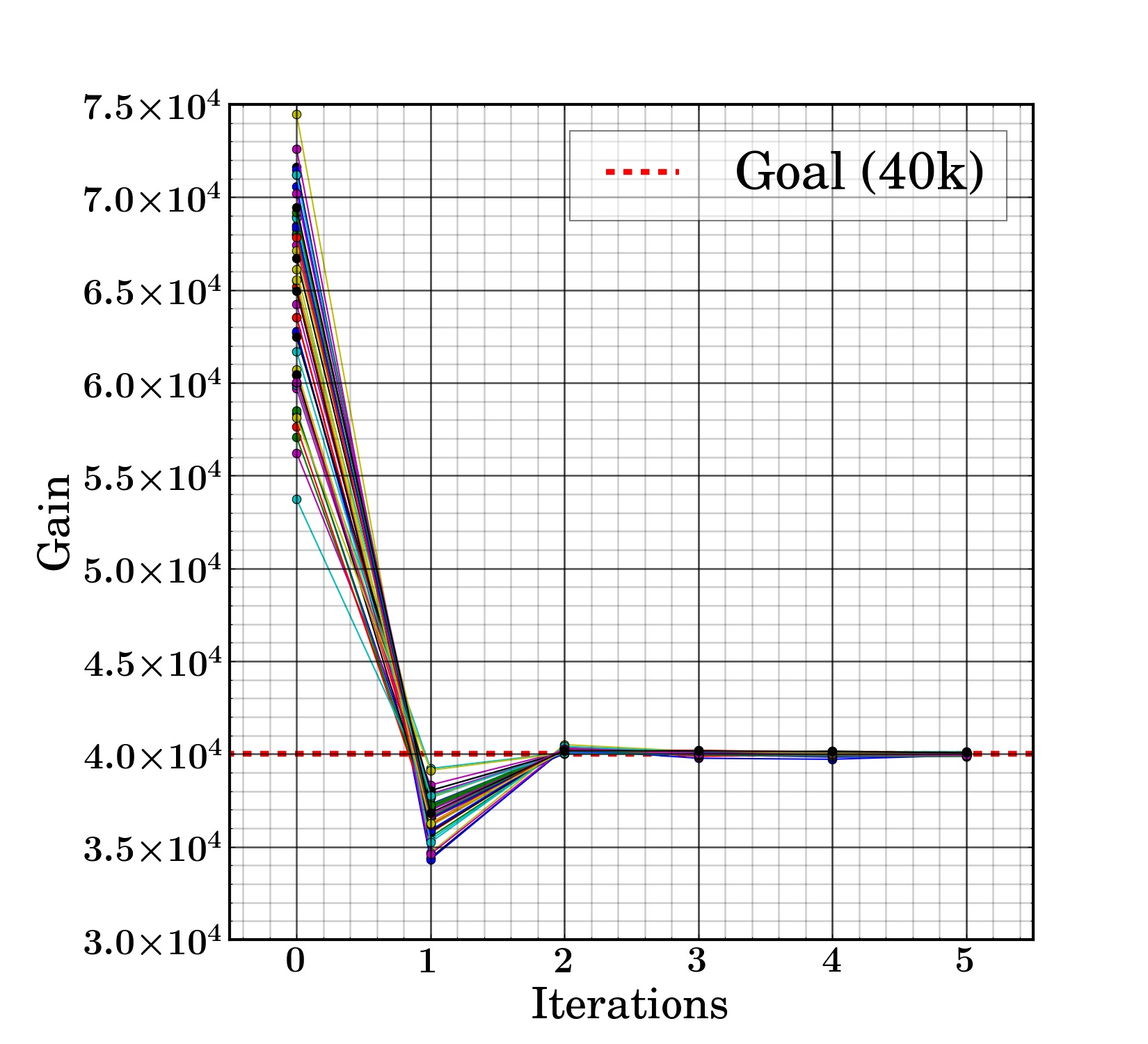}
  \caption{Nominal high voltage tuning iterations for 7 FPMs (49 PMTs). The dashed red line shows the goal value of 40000. The gain estimations are done with the methods described in section \ref{Text:Gain}.}
  \label{FIG:Iterations}
\end{figure}

\begin{figure}[H]
  \centering
  \includegraphics[width=8cm]{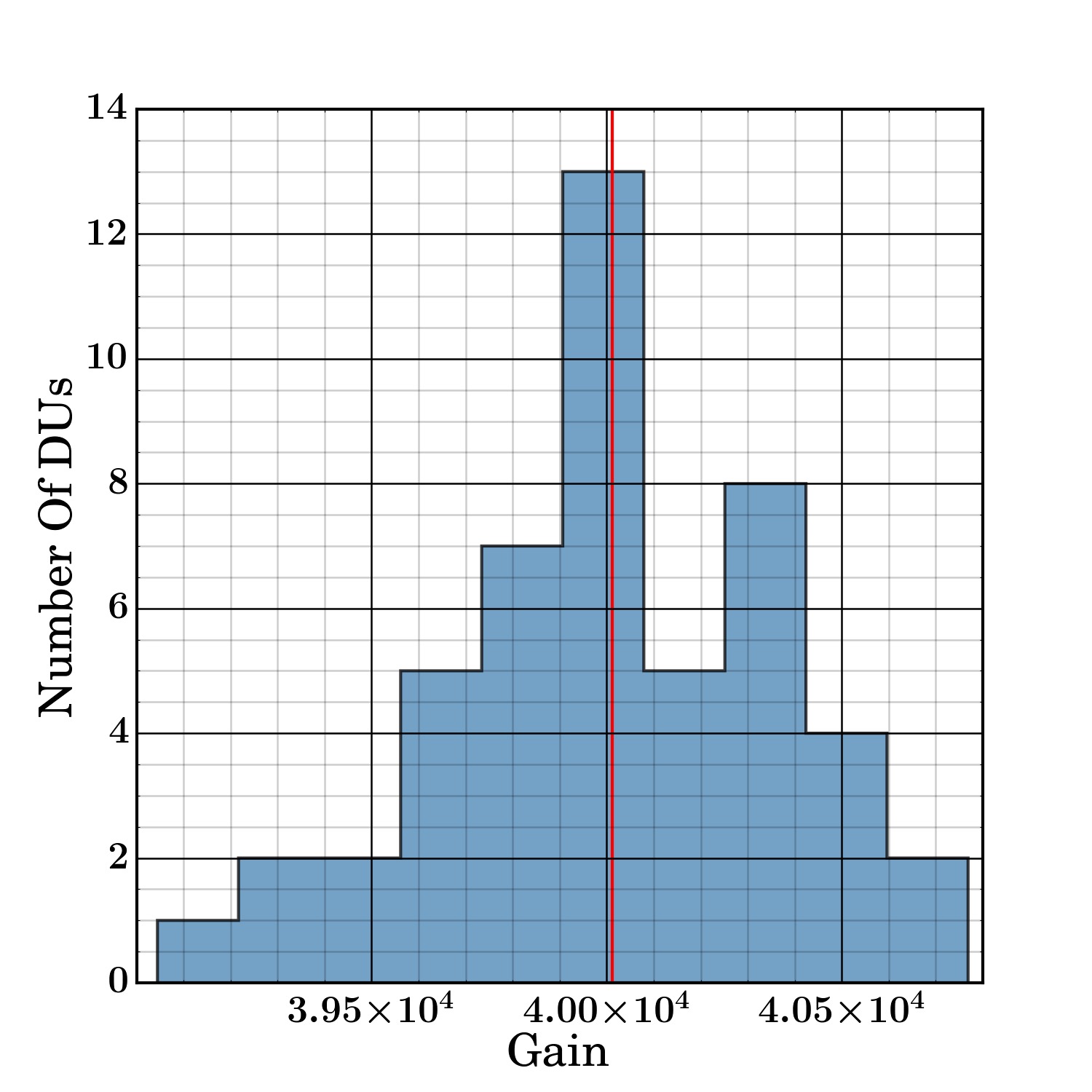}
  \caption{Distribution of the gain at nominal high voltage for 7 FPMs (49 PMTs). The red line shows the mean gain value of 40012. The standard deviation of the distribution is 355.}
  \label{FIG:Gain_hist}
\end{figure}

The nominal high voltage determination method has been developed and extensively tested on 7 FPM prototypes. For a batch of 49 PMTs, the total process duration is about 50~min. Nominal high voltages are reached in less than 5 iterations (see Figure \ref{FIG:Iterations}). Figure \ref{FIG:Gain_hist} shows the histogram of the resulting nominal gains. The mean value of the gain is 40012 and its standard deviation is 355. The standard deviation is $\sim$~0.9\% of the mean. This quantifies the accuracy of the iterative method used to determine the nominal high voltage.

\begin{figure}[H]
  \centering
  \includegraphics[width=8cm]{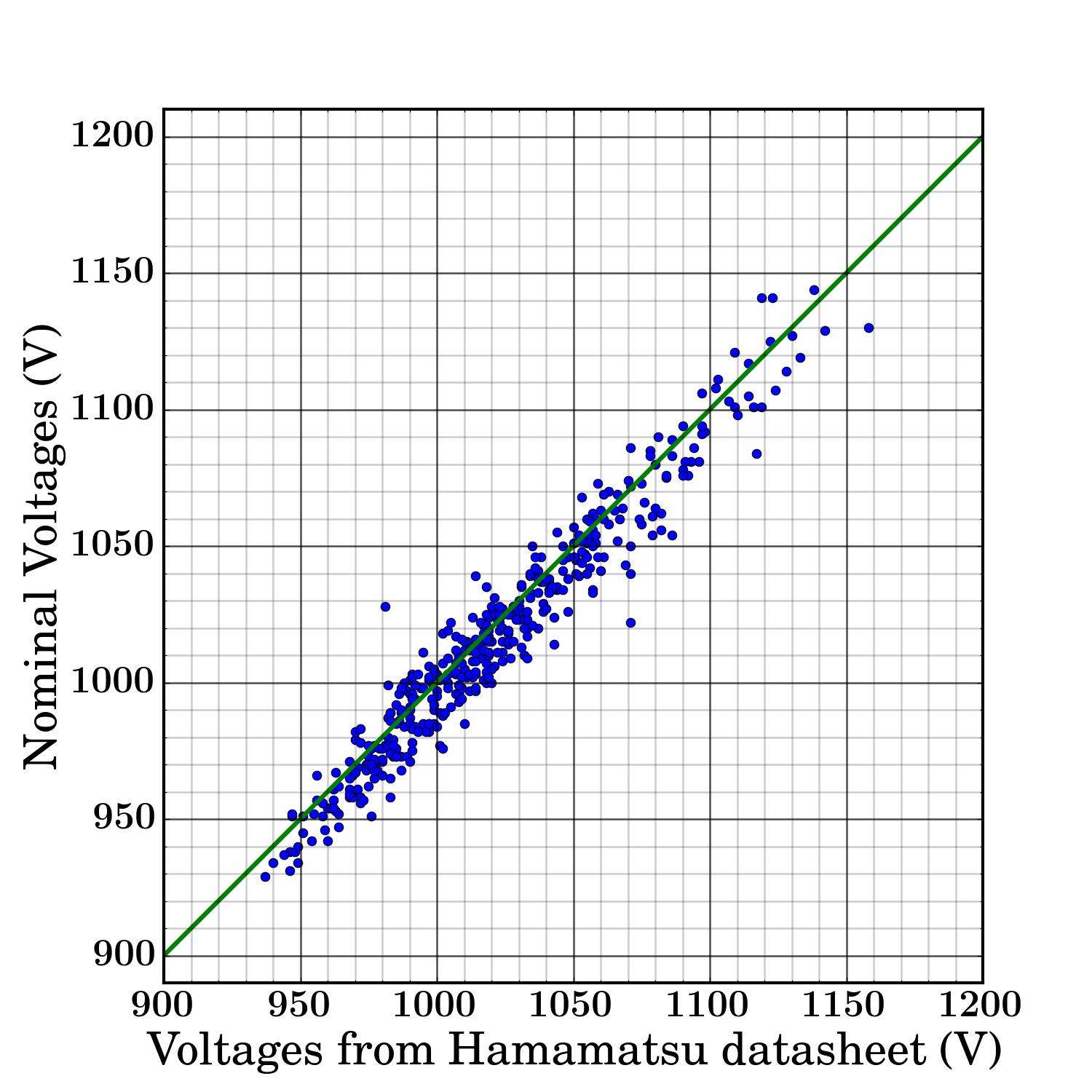}
  \caption{The nominal high voltages for 427 PMTs from our test bench vs nominal high voltages from Hamamatsu data. The green line shows the one-to-one relation between the two voltages.}
  \label{FIG:Nom_vs_Hama}
\end{figure}

The nominal high voltage values of the 427 PMTs measured with our test bench are in agreement with those provided by Hamamatsu (see Figure \ref{FIG:Nom_vs_Hama}), the mean of the differences is 0.5\%, and the standard deviation of the differences is 1\%.
At their nominal gain, the excess noise factor\footnote{This factor is ENF = $\sqrt{1+\sigma_{SPE}^2}$ , see equation 1 of \cite{Mir17}, with $\sigma_{SPE}$ the best fitting single photo-electron rms (i.e. SPE RMS in Figure \ref{FIG:myspefitter_plot}). It describes the effect of the charge dispersion caused by the dynode amplification system.} (ENF) of the 427 PMTs derived from the best fitting photo-electron spectrum models, has a mean value of 1.106 and a standard deviation of 0.005, with minimal and maximal values of 1.096 and 1.131, respectively. This is in good agreement with the previous measurements (ENF ranging from 1.05 to 1.12) made by \cite{Mir17} with the same kind of 7-dynodes PMTs. The mean and the standard deviation of the RMS of the electronics noise pedestal obtained with the best fitting parameters of the photo-electron spectrum model are 20.5\% and 0.6\% of the single photo-electron peak, respectively. The measurements made in the high-intensity regime offer the opportunity to estimate the PMT gain with the photo-statistics (or F-factor) method (see  \cite{FeganReport}, \cite{Bencheikh92}, \cite{Biller95}, \cite{Mir_int_20}) for comparison. This method uses the Poissonian-law based relation between the mean number of photo-electrons and the variance of the measured charges. The gains obtained with the photo-statistics method are similar to the ones obtained with the single photo-electron fitting method; the mean ratio of the photo-statistics gains to the single photo-electron gains is 0.981 with a standard deviation of 0.025.

It has been observed that PMT gain increases with increasing photo-cathode current, which in an IACT is largely determined by the night-sky background (NSB) level, due to the impedances of the last dynodes (e.g. see \cite{Zhong89}). Indeed, at high photo-electron rate, the current in the last stages of the voltage divider circuit is so large that it enhances the potential difference between the last dynodes, leading to an enhancement of the global PMT gain. That effect has been minimised by design, using low-impedance resistors in the last stages of the voltage divider. For NectarCAM we require that the gain variation with NSB rate should be below $2\%$ up to NSB rates of $\sim$~1~GHz, which corresponds to the maximum NSB rate expected during observations at CTA sites.

For the whole batch of 61 FPMs, we did multiple pedestal and high-intensity regime (with 60 ph.e. pulses) acquisitions with NSB rates ranging from $\nu_{nsb}= $ 4~MHz to 15~GHz. For each PMT and each NSB rate, we estimated the gain with the measurement of the mean net charge using the method described above. We estimate the NSB rate from the variation of the pedestal width with illumination. Such a variation is expected due to the single photo-electron pulses induced by NSB photons, which produce fluctuations of the electronic baseline. The NSB rate is derived from pedestal acquisitions with the formula indicated in Eq. \ref{eq:nsb_estimation} where $\sigma_{ped}(\nu_{nsb})$ is the standard deviation of measured charges with a NSB rate $\nu_{nsb}$, $\sigma_{ped}(\nu_{nsb}=0)$ is the standard deviation of measured charges without NSB (i.e. the rms of electronic noise), $G$ is the gain of the PMT, $T$ is the duration of the integration window (in our case we choose $T = 32$~ns). The subtraction of 2~ns from the duration $T$ accounts for time window edge effects since, with a width of $\sim$2.3~ns FWHM (see section \ref{Text:PulseShape}), the anode pulses produced by NSB induced photo-electrons, that start within $\leq$ 2~ns of the end of the time window are not fully integrated. 
\begin{equation}\label{eq:nsb_estimation}
\nu_{nsb} = \frac{\sigma_{ped}^2(\nu_{nsb}) - \sigma_{ped}^2(\nu_{nsb}=0)}{G^2*{\rm ENF}^2*(T-2)}
\end{equation}

\begin{figure}[H]
  \centering
  \includegraphics[width=8cm]{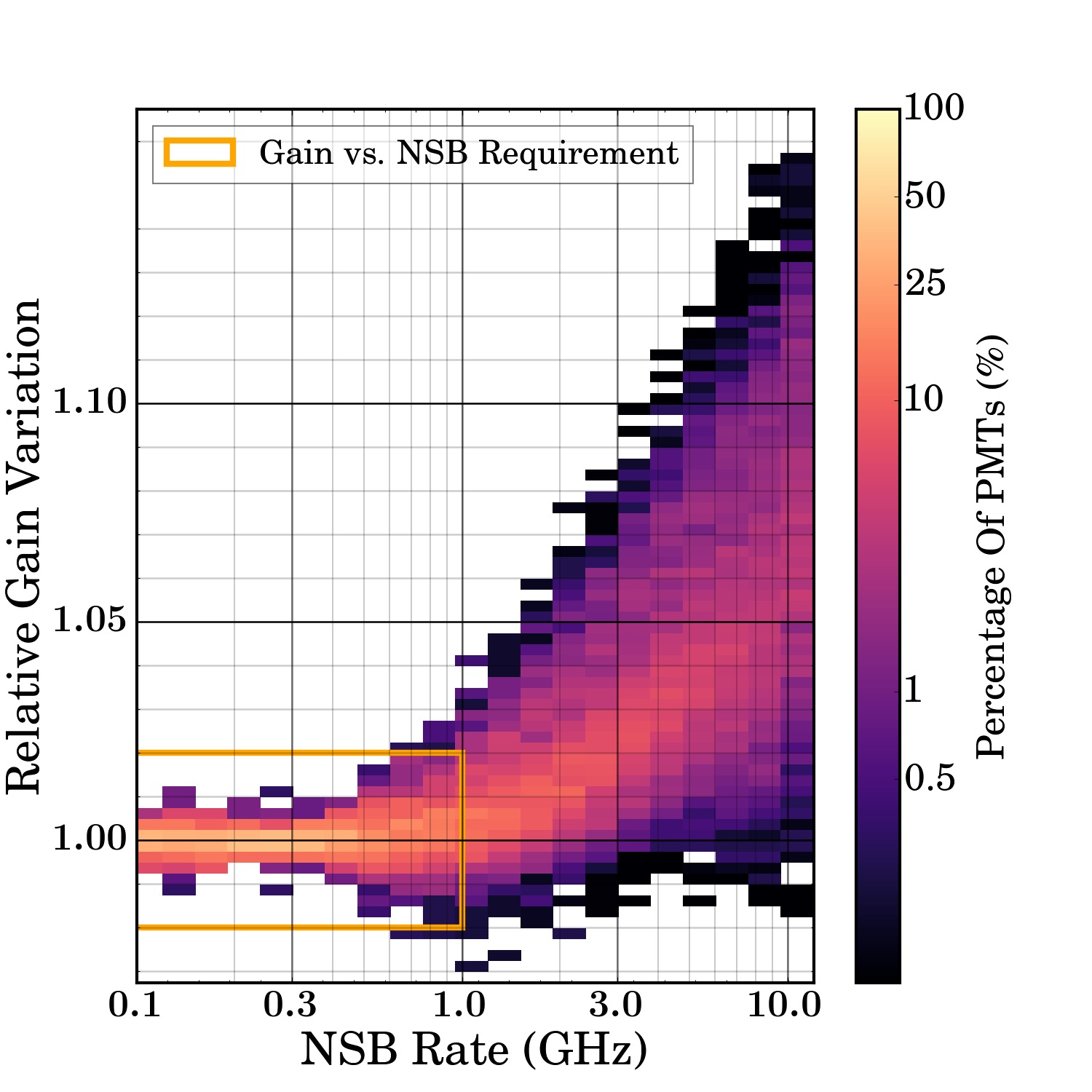}
  \caption{Distribution of the relative gain variation of the 61 FPMs (427 PMTs). The sum of entries in each vertical column is 100\%. The orange line shows the limit on PMT gain variation of 2\%, below a NSB rate of 1~GHz.}
  \label{FIG:NSB_3D_map}
\end{figure}

\noindent
This formula is based on the fact that the variance of the baseline ($\sigma_{ped}^2$) increases with the intensity of random background light (see \cite{Bencheikh92}). Its validity has been verified at low NSB rate by comparing the rates obtained with Eq. \ref{eq:nsb_estimation} with the photo-electron counting rates; the differences between the two rates are $<$~10\% for NSB rates ranging from 2~MHz to 200~MHz.
We compute the relative gain with a reference at $\nu_{nsb}=$~0.3~GHz which is representative of the NSB rate expected during observations with a MST telescope pointing towards the Galactic plane. Figure \ref{FIG:NSB_3D_map} shows the distribution of the relative gain variation versus NSB rate of the 427 PMTs. As expected, the gains increase at high NSB rates. However, below 1~GHz most of the relative gain variations are below 2\%, which is compliant with the NectarCAM requirements.

\subsection{Charge resolution}\label{Text:ChargeRes}
The relative charge resolution is the relative dispersion of the measurements of the number of photo-electrons detected with a PMT when it is illuminated with light pulses of known intensity. The charge resolution is an important characteristic for CTA since it directly influences the final energy resolution for gamma-ray observations. Definitive measurements of the NectarCAM charge resolution will be done at the integration site at IRFU, once the full system is assembled, as this requires a realistic in-camera temperature profile and RFI (Radio-Frequency Interference) environment, which can only be achieved when the full system is assembled and operated in realistic conditions.

In this section, we describe the method we developed to evaluate the charge resolution of the FPMs with our test bench. The charge resolution is measured in the presence of simulated NSB light in order to reproduce some of the conditions expected during observations. The charge resolution tests verify that the signal-to-noise ratio (SNR) is within specification over almost four orders of magnitude of pulse amplitude. This SNR test serves as a "system test" for the FPM, verifying the integrity of the signal chain, and that there are no unexpected sources of noise present (pickup from the CW supply, ringing in the pre-amplifier, etc). In the production phase, failure of these SNR tests would require the manufacturer to diagnose and repair the faulty DU or FPM boards.

The charge resolution is measured with PMT high voltages set to their nominal values (i.e. a gain of 40000 -- see section \ref{Text:Gain}) and with the NSB simulation system tuned to produce a rate of $\sim$~0.125~ph.e./ns (i.e. a NSB rate of $\sim$~125~MHz\footnote{Actually, the NSB rates measured with our PMTs had a mean of 124~MHz with a standard deviation of 15~MHz depending on their position on the FPM test bench.}), as specified by CTA (\cite{HintonMSTReq14}). During a data taking run, the PMTs are illuminated with 4000 light pulses at a given intensity emitted by the nano LED. Several runs are performed with light pulse intensities ranging from $\sim$~1 to $\sim$~5000 ph.e. per light pulse. An additional acquisition is made without light pulses to estimate the pedestal characteristics and to control the NSB rate. For each set of 4000 events, we extract the charges by integrating the anode pulses in a 16~ns time window and calculate the mean net charge value $\bar{Q}$ and the standard deviation of the distribution $\sigma_Q$. The charge resolution is then given by $r = \nicefrac{\sigma_Q}{\bar{Q}}$.
At nominal PMT gain, the anode signal of the HG line is near saturation when $\bar{Q} \gtrsim$~160 ph.e. Consequently, for larger pulses, the charge is measured with the anode signal from the LG line. In the overlap range when the anode signal of the LG line is significant enough and the one of the HG line does not saturate (from $\bar{Q}$ = 70 to $\bar{Q}$ = 150 ph.e.), we measured an average HG/LG mean net charge ratio of $\langle r_{hg/lg} \rangle$ = 13.58 with a standard deviation of 0.29 for all the data of the 427 PMTs. As expected, this mean net charge ratio is in quite good agreement with the ratio of the PACTA gains (see section \ref{Text:FPM_Nect}).

\begin{figure}[H]
  \centering
   \includegraphics[width=8cm]{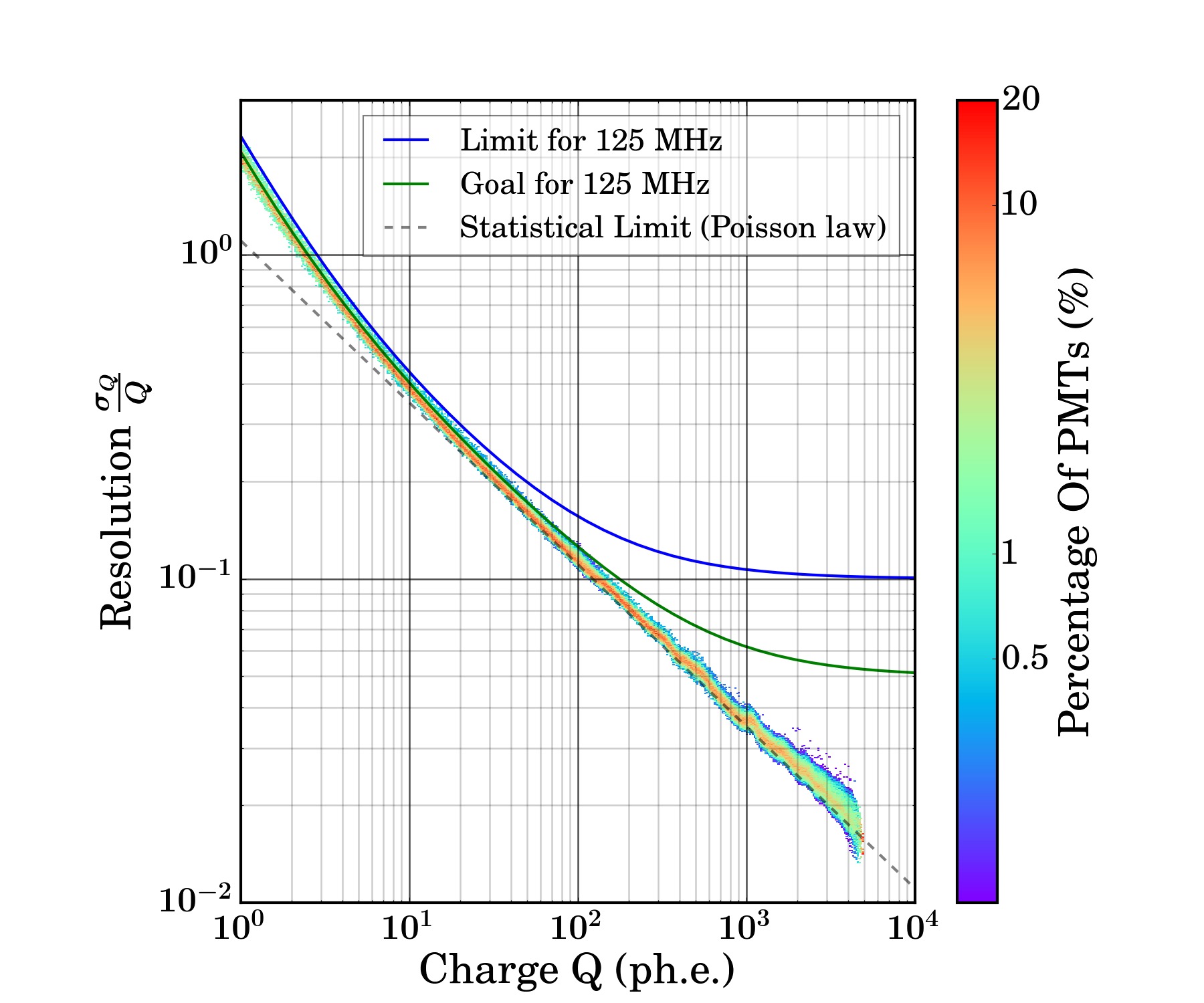}
  \caption{Charge resolution distribution measured with a NSB rate of $\sim$~125~MHz for 61 FPMs (427 PMTs). The sum of entries in each vertical column is 100\%}
  \label{FIG:Charge_res_3D_map}
\end{figure}

Figure \ref{FIG:Charge_res_3D_map} shows the distribution of the charge resolutions measured with the 427 PMTs as a function of the mean net charge of the pulse $\bar{N}_{pe}$ expressed in photo-electron unit; i.e. $\bar{N}_{pe} = \bar{Q} / \bar{Q}_{SPE}$ for charges measured with the HG line and $\bar{N}_{pe} = \bar{Q} / \bar{Q}_{SPE} \times r_{hg/lg}$ for charges measured with the LG line. The dashed curve is the theoretical lower limit corresponding to the Poissonian statistics ($\nicefrac{1}{\sqrt{\bar{N}_{pe}}}$), broadened by the ENF of the PMT. The blue curve shows the limit on charge resolution for acceptance of the DU/FPM for use in NectarCAM for a NSB rate of 125 MHz, while the green curve shows the goal for a NSB rate of 125 MHz. The measured charge resolutions for all of the 427 DUs are within the acceptable bounds. As expected, for high mean net charges (above 100 ph.e.), the measured charge resolution asymptotically tends toward the Poissonian limit broadened by the ENF because, in that case, the main source of fluctuations is the number of photo-electrons, well above other sources such as the electronic noise or NSB. This is not the case for low mean net charges whose fluctuations are due to a combination of Poissonian, electronic noise and NSB fluctuations. The drop of the charge resolution for charges $\gtrsim$~4000~ph.e. (see Figure \ref{FIG:Charge_res_3D_map}) is not an improvement of the charge resolution. It originates from an underestimation of charges due to the saturation of the anode signal in the LG line. Indeed, the electronics chain (PACTA, ACTA and ADC) was designed to process anode current pulses up to $\sim$ 3000 ph.e. when the PMT gain is 40000 (see section \ref{Text:FPM_Nect}).
The presented charge resolution does not take into account the possible systematic effects that could deteriorate the overall charge resolution when the PMTs are installed in the camera, such as e.g. the variations of the pedestal levels and PMT gains with temperature, which lead to inaccurate estimation of charges. The impact of those systematic effects on the charge resolution at the camera level will be presented elsewhere, in a dedicated publication.

\subsection{Pulse shape}\label{Text:PulseShape}

The first level of trigger (so called level 0) of the NectarCAM electronics is set to the level of $\sim$~5 ph.e. amplitude of the anode signal. The timing performance of this trigger is directly related to the impulse response function of the system for a single photo-electron. Due to the limited bandwidth of the detection chain, the impulse response function is Gaussian-like with a finite rise time and a delay (transit time) after the photo-electron generation. The transit time is affected by fluctuations that are determined by the point of illumination on the photocathode and the propagation of the electron cascade within the PMT. In this section, we present measurements of the single photo-electron pulse shape as well as its timing properties.

\begin{figure}[H]
  \centering
  \includegraphics[width=8cm]{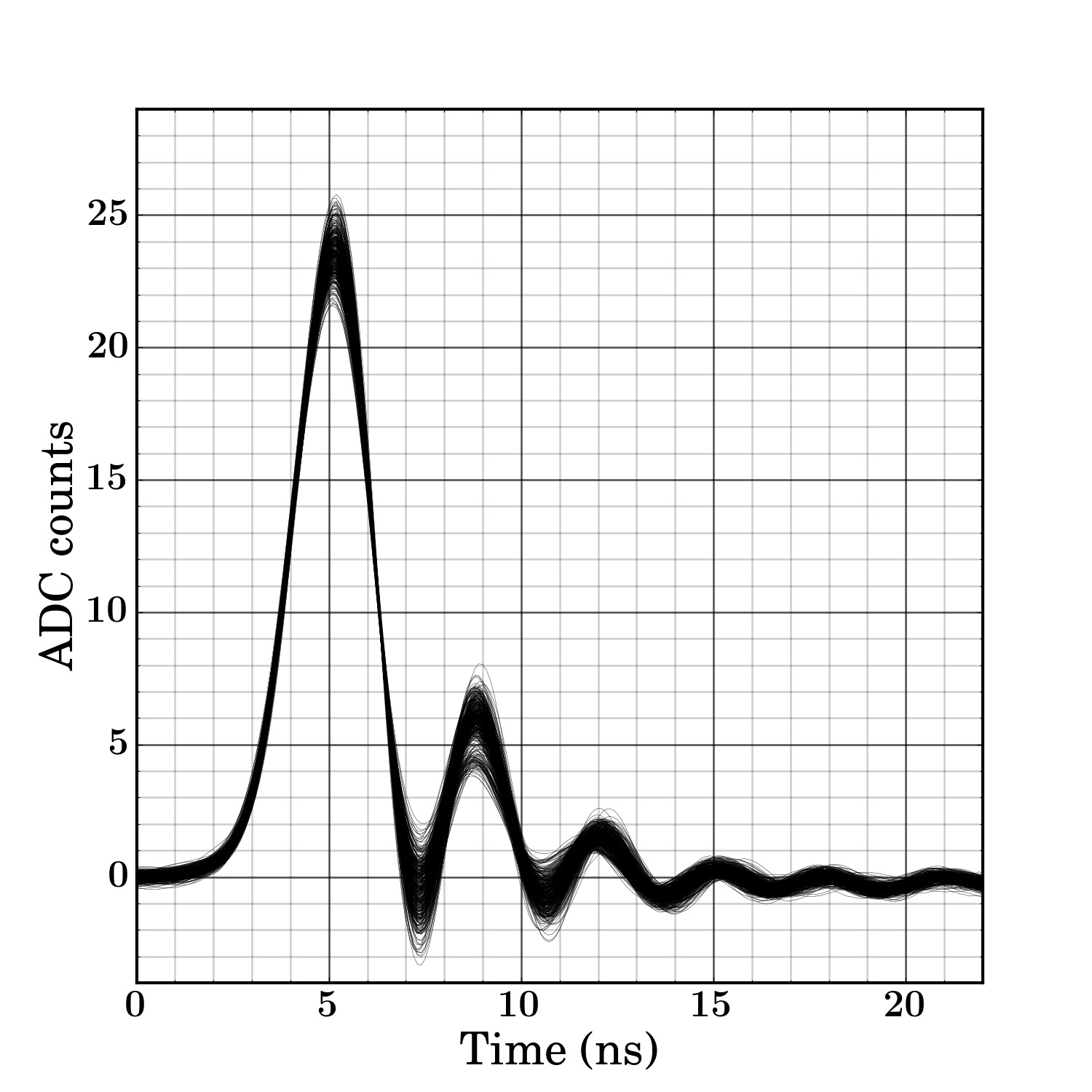}
  \caption{Mean single photo-electron pulses of the 427 PMTs measured with a gain of 40000. One curve is plotted for each PMT.}
  \label{FIG:Stacked_pulses}
\end{figure}

\begin{figure}[H]
  \centering
  \includegraphics[width=8cm]{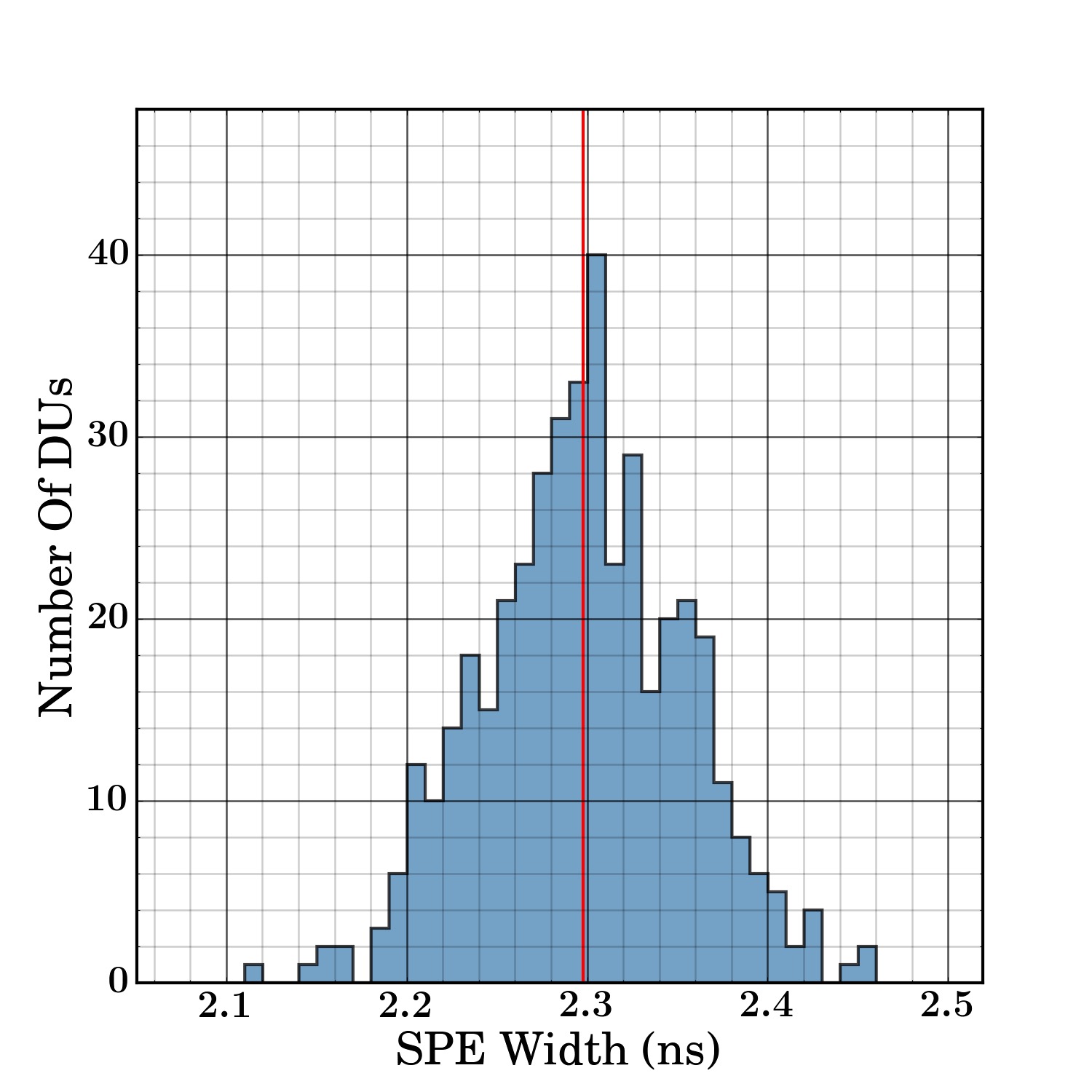}
  \caption{Distribution of the single photo-electron pulse width of the 427 PMTs at a gain of 40000. The average and standard deviation of the distribution are 2.29 ns and 0.055 ns, respectively.}
  \label{FIG:FWHM_pulses}
\end{figure}

\begin{figure}[H]
  \centering
  \includegraphics[width=8cm]{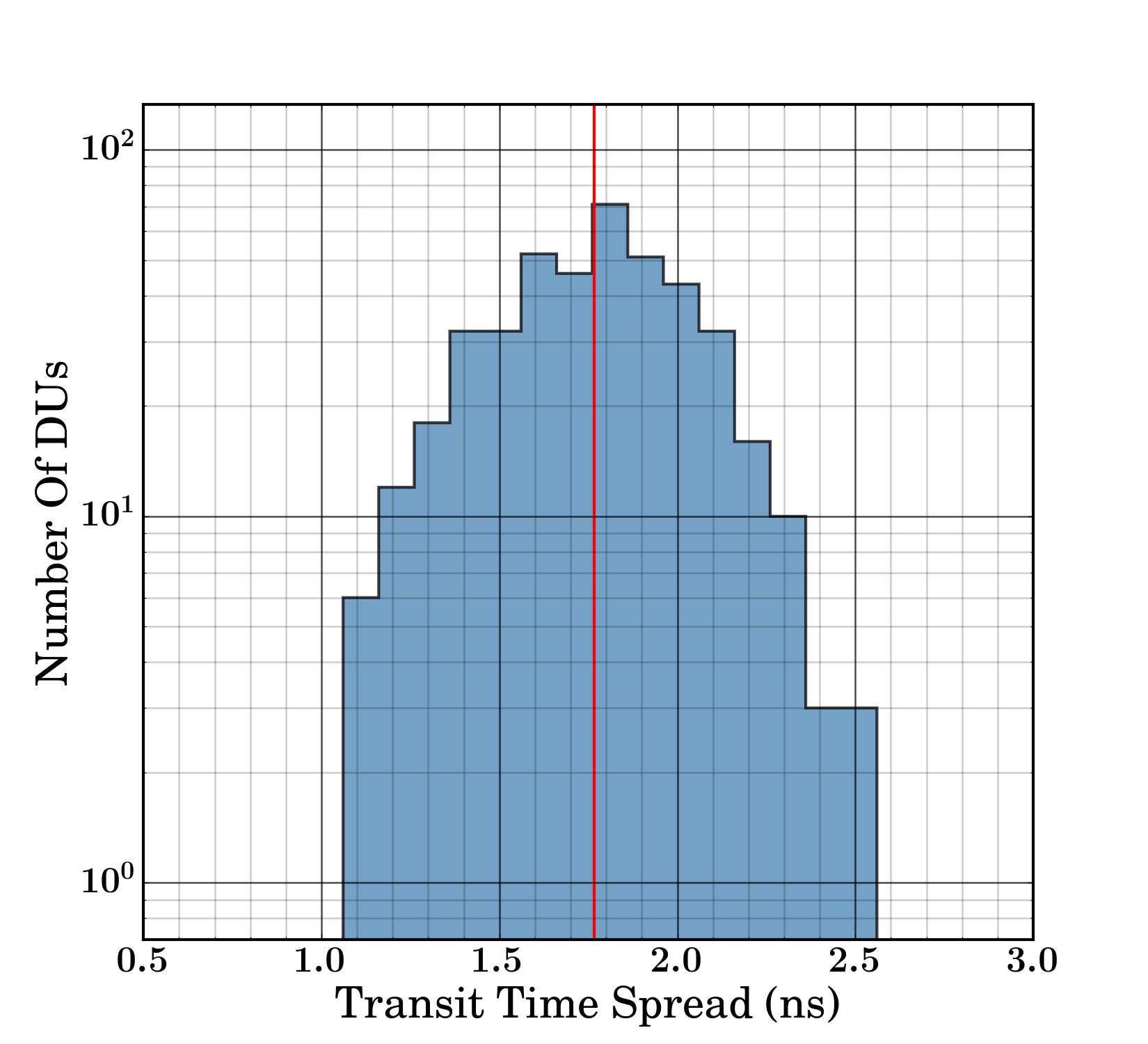}
  \caption{Distribution of the TTS values measured with the 427 PMTs at a gain of 40000. The average and standard deviation of the distribution are 1.76 ns (red line) and 0.28 ns, respectively.}
  \label{FIG:TTS_distribution}
\end{figure}

We measured the single photo-electron pulse shapes of the 427 PMTs with the data acquired for the single photo-electron measurements at nominal gain (see section \ref{Text:Gain}). In order to avoid events from pedestal or low-charge components and events produced by two photo-electrons, we selected pulses with net charges $Q$ of $\pm$~2 ADC counts around the single photo-electron peak $\bar{Q}_{peak}$ of the photo-electron spectrum: i.e. $Q \in [\bar{Q}_{peak} - 2, \bar{Q}_{peak} + 2]$ with $\bar{Q}_{peak} \simeq1.14 \cdot\bar{Q}_{SPE} \simeq 66$ ADC counts. To get a better timing resolution in order to determine the mean pulse shape from the acquisition, we re-binned each pulse from the sampled resolution of 1~ns to a resolution of 0.1~ns by quadratically interpolating the measured values. Before the quadratic interpolation, the arrival time of each pulse is estimated by fitting each measured pulse with an analytical model composed of three Gaussian functions, where the first Gaussian accommodates the main pulse, and the two other Gaussians model oscillations that occur in the trailing wing of the pulse. The position of the first Gaussian is used to estimate the arrival time of the pulse within the sampling window. The arrival time is the time delay with respect to the trigger signal from the Flash LED pulser. The interpolated pulses, with their sampled resolution of 0.1~ns, are shifted by their respective time delays. The resulting shifted pulses are stacked together and then averaged. Figure \ref{FIG:Stacked_pulses} shows the resulting averaged single photo-electron pulses of the 427 PMTs. The shape of mean pulses is in agreement\footnote{The delays of the second and third pulse peaks, as well as the ratio of the amplitude of the main pulse peak to the amplitude of the second pulse peak are similar.} with that measured directly at the output of the PACTA with an oscilloscope (internal NectarCAM note \cite{Sanuy_private20}). The pulse is characterized by a first impulse with an amplitude of about 23 ADC counts, followed by damped oscillations with a frequency of about 280 MHz. The total duration of the pulse is about 12~ns when the second oscillation is accounted for. Figure \ref{FIG:FWHM_pulses} presents the distribution of the single-photo-electron pulse widths (FWHM) of the 427 PMTs measured with the extracted pulse shapes of Figure \ref{FIG:Stacked_pulses}. The average width of the primary pulse is 2.29~ns FWHM with a standard deviation of 0.05~ns. This value is smaller than the mean value of $\simeq$~3~ns measured by \cite{Toyama15} and the value of 2.86~ns with a standard deviation of 0.05~ns provided by Hamamatsu which can be explained by the fact that we average the pulses after charge selection and time delay correction. The LST team measured a single photoelectron pulse width of 2.6~ns FWHM by minimizing the pulse broadening due to the dispersion in time delay with a mask which has small pinholes in front of the photocathodes of 8-dynode PMTs (see \cite{LSTmeeting2020}). Without selection and time delay correction we measure an average pulse width of 3.09~ns with a standard deviation of 0.10~ns, in agreement with \cite{Toyama15} and the values measured by Hamamatsu. The dispersion in time delays is attributed to differences in the electron path lengths within the PMT for different incident photon locations on the photocathode. Electrons generated at the center of the photocathode have different trajectories to the first dynode than electrons generated at the outer parts. The time delay dispersion is characterized by the transit time spread (TTS) which is the FWHM of the time delay distribution. Figure \ref{FIG:TTS_distribution} shows the distribution of the TTS values measured with the 427 PMTs. We measure TTS values with a mean of 1.76~ns and a standard deviation of 0.28~ns, in agreement with (1.81 $\pm$ 0.12)~ns measured by \cite{LSTmeeting2020} and with the Hamamatsu technical notes ($<$~2~ns) but not with the initial requested specifications for the PMTs of CTA ($<$~1.5~ns FWHM, see \cite{Toyama15}).

\subsection{Afterpulses}\label{Text:AfterPulses}

Afterpulses are produced by residual ions within the PMT that can release up to $\sim$~50 ph.e. when they interact with the photocathode or with a dynode. They degrade the performance of the Cherenkov camera in several ways. During observations, these spurious pulses mimic the signal of Cherenkov photons and may increase the electronics dead-time of a camera due to an increased trigger and read-out rate. If their rate is too high the trigger threshold level has to be increased to reduce the dead-time, resulting in an increase of the energy threshold of gamma rays; afterpulses have a direct impact on the energy threshold of a Cherenkov camera (see \cite{Mir97}). Moreover, afterpulses affect the separation of events due to gamma from the (background) events due to hadrons in the shower image analyses. The PMTs were designed by Hamamatsu, in collaboration with CTA, to have an afterpulse probability of $<$~2~$\times$ 10$^{-4}$ when the pulse threshold is set to 4 ph.e. (see \cite{Toyama15}). Here we verify that the PMTs furnished to NectarCAM meet this specification.

\begin{figure}[H]
  \centering
  \includegraphics[width=8cm]{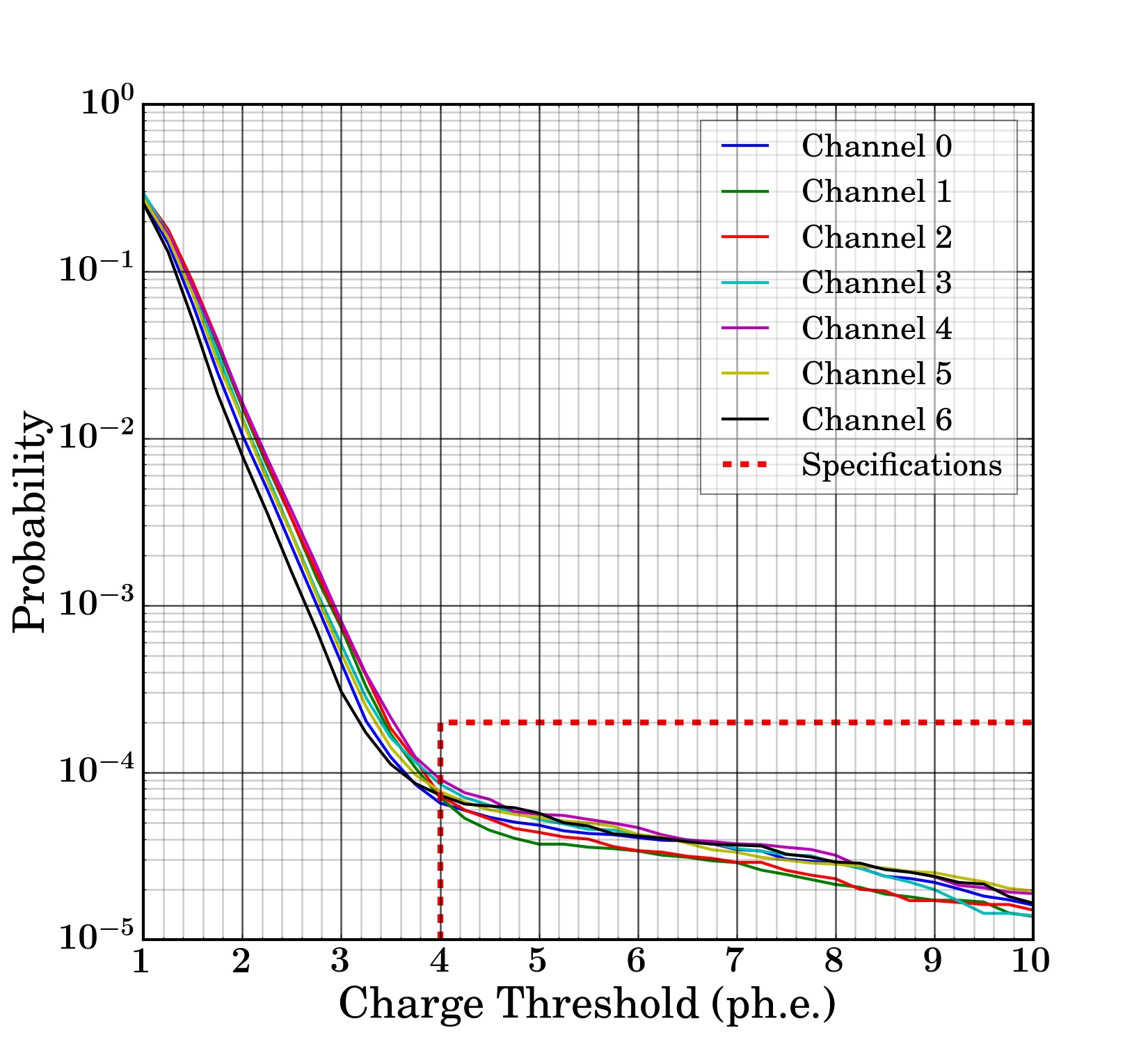}
  \caption{Afterpulse probabilities for the 7 PMTs of a focal plane module prototype measured at a gain of 100k for comparison with Hamamatsu measurements. The dashed line shows the Hamamatsu specifications. The probability distribution is dominated by NSB photons below $\sim$ 3.5 ph.e.}
  \label{FIG:Afterpulse_probability}
\end{figure}

With the FPM test bench, the afterpulse rate was measured by illuminating a FPM with a continuous rate of simulated NSB photons. The simulated NSB photons produce afterpulses which might occur by chance in the acquisition windows. The distribution of the net charges measured in the acquisition windows allows highlighting the contribution of afterpulses since the accelerated residual ions produce on average more photo-electrons per interaction than the NSB photons. The NSB rate was set to a value that allows for a clear detection of individual photon pulses from the NSB and the characterization of their waveforms, while producing enough afterpulses. The acquisition system was triggered at a frequency of 800 Hz. We used PMT gains of 100000, which is the same setting that is also used by Hamamatsu to determine the afterpulse probabilities. Adoption of such a gain improves the pulse detection and the accuracy of the amplitude and net charge measurements. 
The number of afterpulses is obtained by counting the number of pulses with amplitude and net charge larger than a given number of photo-electrons. 
To estimate the afterpulse probability, the resulting number counts are divided by the total number of photo-electrons which is derived from the measured NSB rate. Figure \ref{FIG:Afterpulse_probability} presents the event probability as a function of the threshold net charge in units of photo-electrons, obtained from 1.5 $\times$ 10$^6$ waveforms with a duration of 48~ns each \footnote{48 ns is the maximum duration of an acquisition with our test bench.} and a mean NSB rate of $\sim$~20 MHz. The shape of the distribution shows the contributions of NSB pulses and afterpulses in the event counts. The probability distribution is dominated by NSB pulses in the low charge range (up to $\sim$ 3.5 ph.e.) and by afterpulses in the high charge range. The afterpulse probability ranges from 0.6 $\times$ 10$^{-4}$ to 0.9 $\times$ 10$^{-4}$ for charges $>$ 4 ph.e., which is in agreement with the PMT specifications and previous measurements (see \cite{Toyama15}, \cite{Mir17} and \cite{Mir16}). 

\section[Conclusions]{Conclusions}

We developed a test bench to measure the performance of NectarCAM focal plane instrumentation. The test bench is fully functional and the measurements, performed with 61~FPMs (427 PMTs), show that the performance of the focal plane modules is in agreement with our expectations, and with measurements done previously on individual PMTs. Our tests validate the design of the FPMs for use with NectarCAM and CTA. For each PMT we find high-voltage values producing a nominal gain of 40000 that are similar to those determined by Hamamatsu. We measured excess-noise factors with a mean of 1.106 with a standard deviation of 0.005. At nominal gain, the dark pedestal RMS values of $\sim$~20\% of the photo-electron peak are compliant with NectarCAM requirements. The gains are stable within $\pm$~2$\%$ when PMTs are subject to NSB rates ranging from a few MHz up to $\sim$~1~GHz. The charge resolutions measured from 1 to $\sim$~3000~ph.e., with a NSB rate of 125~MHz, are within specification. The mean single photo-electron pulse widths and transit time spreads are 3.09~ns with a standard deviation of 0.10~ns and 1.76~ns with a standard deviation of 0.28~ns, respectively, consistent with the values measured by \cite{HamamatsuPMTDataSheet}, \cite{Toyama15}, \cite{Mir17} and \cite{Mir16} . After correction of the time delay with respect to the trigger, we obtained a mean single photo-electron pulse intrinsic width of 2.29~ns with a standard deviation of 0.05~ns. The afterpulse rates measured with 7~PMTs are $< 2\times10^{-4}$ for charges $>$ 4 photo-electrons, in accordance with the specification of Hamamatsu.

We calculated the autocorrelation functions of single photoelectron pulses measured with 427 PMTs of the Focal Plane Modules for NectarCAM. After removing the contribution of noise fluctuations, the width of the auto-correlated single photoelectron pulses are of $\sim$~3.2 ns in average. The bandwidths of the full acquisition chain, calculated with the Fast Fourier Transform of auto-correlated pulses, are about $\sim$~77 MHz. The average bandwidth of PMTs alone is $\sim$~82 MHz, after correction of the electronic chain (PACTA, ACTA and Nectar chip) response. The slight difference between those values shows that the electronic chain does not reduce significantly the bandwidth of PMTs; the electronic chain response is adapted to the response of PMTs. The similarity between the average auto-correlated single photoelectron pulses and the auto-correlation of the extracted mean single photoelectron pulses suggests that the method to extract the single photoelectron pulse does not twist too much the shape of single photoelectron pulse.

After validation with our test bench, the 61~FPMs were installed in a NectarCAM prototype camera which was mounted on a MST telescope prototype (see \cite{Garczarczyk18}) at Adlershof (Germany) during May--June 2019 to perform measurements and tests under realistic operation conditions and to record images of air showers (e.g. see Figure~4 of \cite{Glicenstein19}). The results of that observation campaign will be presented elsewhere (\cite{NCam20}). 
We upgraded our test bench to allow the characterization of the 265 FPMs that are produced by industry to furbish the NectarCAM Qualification Model (QM), which fully reflects the design of the final model. The 61 FPMs are spare parts of the QM. One of the improvements consists of replacing the wooden shelf on which FPMs are installed up to now (see Figure \ref{FIG:FPM_TB_pic}) by a robust mechanical structure. That structure is also designed to enhance the ergonomy for easy and reliable plugging and unplugging of FPMs by the test operator. With regards to measurements and analyses, we have identified a need to improve the method for estimating the PMT gains. So far we estimate the gains by assuming a low-charge component fraction of 0.17 per single photo-electron for each PMT but our analysis suggests that this fraction may vary significantly among PMTs. This may result in systematic uncertainties of about 1.3$\%$ on the gain (see \cite{Caroff19}). In order to reduce these uncertainties, we will implement a new method in our test bench, where we will estimate the low-charge component from single photo-electron measurements at high gain. 

During the mass production of cameras for CTA, the FPMs of NectarCAM will be produced in industry including a characterization of the modules by the manufacturer using a dedicated test bench before delivery. The fully automated test bench will perform a series of tests per batch of 7 FPMs. That test bench will comprise the same elements as the FPM test bench described in this paper but it will be mounted on a mechanical structure consisting of a dark box with a hinged lid so that it can be used within a production chain during daylight. Moreover, the measurements that will be performed in industry will be optimized to reduce time and cost. In that way, fully qualified and characterized FPMs for the NectarCAM can be delivered by an industrial partner, avoiding the need for costly performance checks at the lab and guaranteeing the delivery of homogeneous, high-quality hardware for NectarCAM on CTA. 

\section*{Acknowledgement}
The authors are grateful for the support of the R\'egion Occitanie and Microtec. The authors want to thank R. Mirzoyan and D. Nakajima for useful discussions. This work was conducted in the context of the NectarCAM Project of CTA. The CTA Consortium gratefully acknowledge financial support from the following agencies and organizations listed on this webpage : https://www.cta-observatory.org/consortium\_acknowledgments/ .

\medskip
\noindent This paper has gone through an internal review by the CTA Consortium.

\printcredits

\footnotesize

\vskip12pt

\makeatletter

\def\pct{\expandafter\@gobble\string\%}

\immediate\write\@auxout{\pct\space This is a test line.\pct }

\onecolumn
\end{document}